# On the impossibility of isothermal heat transfer and its implications for thermal physics

A. Paglietti

University of Cagliari, 09123 Cagliari, Italy
E-mail: paglietti@unica.it

**Abstract.** The physical impossibility of heat transfer under isothermal conditions implies that the classical expression for the entropy of the ideal gas may not be compatible with the internal energy of the gas itself. A corrected expression of the ideal gas entropy is derived here. It is independent on volume. This result is shown to be at a variance with the statistical interpretations of entropy as a quantity that is related to the number of microstates compatible with the macroscopic state of the system. It also offers a better understanding of the thermodynamic notion of entropy. The present analysis also establishes a general equation that links the entropy of a system to its internal energy.

# 1. Introduction

At uniform constant temperature, no heat exchange can occur. This basic fact of thermodynamics follows from the very definition of heat as *that which is transferred between two systems by virtue of temperature difference only* ([1], p. 73). The heat defined in this way is measured by calorimetry. The equivalence of heat and energy is, in a sense, a secondary issue, since it results from additional experimental evidence. An obvious consequence of the thermodynamic definition of heat is that no heat can flow between any two systems in thermal equilibrium if the temperature of both systems remains constant. In particular, the heat absorbed or lost by a system in a process in thermal equilibrium with its surroundings must vanish identically.

Fundamental as they are, the above considerations do not appear to have been exploited in full. The present paper shows that many important issues of thermodynamics, especially (but not exclusively) those concerning entropy, have not taken into account the fact that the isothermal transfer of heat is impossible. As a result, the very expression for the entropy of an ideal gas that is currently in use turns out to be incorrect. The consequences of this are particularly critical in view of the role played by the entropy of an ideal gas in physics.

The standard procedure for determining the entropy of an ideal gas is reviewed in the next section. Although mathematically correct, this procedure does not take into account the physical fact that in the absence of temperature differences, no heat transfer can occur between the gas and its surroundings. The physically correct entropy expression is obtained, and it is shown that the ideal gas entropy cannot depend on volume.

An alternative proof of the same result is given in Sections 3 and 4. The analysis presented here is general and can be applied to any thermodynamic system, and in particular to ideal gases. In Section 3, we first decompose the internal energy of a system into two parts: a temperature-independent component and a temperature-dependent one. The latter is referred to as *intrinsic heat*, since it represents the amount of heat that the system would release if cooled down to a temperature of absolute zero while keeping constant all the state variables other than temperature. We then establish the relationship between intrinsic heat and entropy.

From this result, we derive, in Section 4, an equation that relates the entropy of a system to the internal energy of the system. This equation enables us to determine the entropy of a system from the internal energy. By applying it to the internal energy of an ideal gas, we recover the same expression for the ideal gas entropy that we found in Section 2, thus confirming this result.

A consequence of the impossibility of isothermal heat transfer is that any isothermal process in thermal equilibrium with the surroundings must also be adiabatic. As evidenced in Section 5, this implies that all the reversible isothermal processes of an ideal gas must be workless. In the same section, we show that the classical formulae for the work and heat supplied by an ideal gas via a reversible isothermal process can in fact be ascribed to a different kind of process, referred to as pseudo-isothermal process. These processes can approximate a strictly isothermal process to any degree of accuracy, and do not stand in contradiction to the thermodynamic notion of heat.

In Section 6, we prove that the classical expression for the ideal gas entropy is also inconsistent with the third law of thermodynamics, thus further supporting the validity of the present analysis. We close the section by showing that the free expansion of an ideal gas in a vacuum is a thermodynamically reversible process. Finally, a reversible way of recovering the initial state of the gas is illustrated.

The implications of the present results for statistical mechanics are considered in Section 7. The involvement of the ideal gas entropy in this theory originated from the widespread belief that it could provide an explanation of thermodynamics in terms of classical mechanics. As described by Mehra [2] in the first volume of his monumental work, this belief was firmly rooted in the minds of many great scientists in the late 1800s and the first two decades of the 1900s. Their efforts eventually led to the linking of entropy to the number of available microstates of a system, as summarised by Boltzmann's well-known entropy equation and its subsequent variants. The role of the classical expression of the entropy of the ideal gases was essential to this result. Even at that time, however, that linking did not receive unanimous consensus and it is still often questioned today. References [3]-[6] and the papers quoted therein contain a reasonably comprehensive exposition of the various objections raised against this interpretation.

The present paper, however, is independent of these objections. Its concern is classical thermodynamics, the correct expression of the entropy of the ideal gases, and the consequences of any correction to this expression for other fields of physics.

Finally, Section 8 shows why the experimental determination of the entropy change of a system by means of processes at constant pressure yields the same results irrespectively of whether the entropy of the considered ssystem depends on volume or not.

Essential to the analysis that of the present paper are the concepts of intrinsic heat, simple heating/cooling processes, and maximum absorbable heat. The author introduced these concepts in [11]. The same concepts were reviewed and applied in some comparatively recent studies on the application of thermodynamics to fluid mechanics, electrochemistry and biological physics [12]-[15]. These works, however, did not focus on the primary issue of thermodynamics considered in this paper. For this reason, and to make the present paper self-contained, the same concepts are presented here in a more organic and practical way.

## 2. Assessment of the classical procedure for determining the thermodynamic entropy of an ideal gas

Let us refer to a homogeneous system at a uniform temperature. For this system, the first and second laws of thermodynamics can be written in differential form as

$$dU = dW + dQ \tag{1}$$

and

$$dS \geq \frac{dQ}{T}, \tag{2}$$

respectively. Here, $U$ is the internal energy of the system, $W$ is the work done *on* it, $Q$ is the heat absorbed, and $S$ and $T$ are the entropy and the absolute temperature of the system. The equality sign in expression (2) applies to reversible processes, i.e., quasi-static processes that are in thermodynamic equilibrium with the surroundings and whose direction can be reversed by an *infinitesimal* change in the external conditions (cf. e.g. [13], p. 22 and [14] p. 18). To these processes, therefore, the equation

$$dQ = T\,dS \tag{3}$$

applies. As apparent from this equation, the second law of thermodynamics requires, in particular, that no change in the system's entropy can occur in a reversible process if the system does not exchange heat with the surrounding world.

The above equations are among the most basic equations on which classical thermodynamics is based. The entropy appearing there is, therefore, the thermodynamic entropy. It is defined as the state function whose change for a reversible process from state A to state B is given by:

$$S_B - S_A = \int_A^B \left.\frac{dQ}{T}\right|_{\text{rev}}. \tag{4}$$

This equation reduces to Eq. (3) when states A and B are in the same infinitesimal neighborhood and a reversible process joining them is considered.

By introducing this equation into the energy balance law (1), we obtain

$$T \, dS = dU - dW. \tag{5}$$

This equation is often referred to as the fundamental thermodynamic relation (cf. e.g. [15], p.153 or [16], p.3), although it applies only to reversible processes.

To define the state of *n* moles of an ideal gas, we use the volume, *V*, and temperature, *T*, of the gas. In terms of these variables, the internal energy of an ideal gas is given by

$$U = U(T) = n \, c_V \, T, \tag{6}$$

to within an inessential constant, which can safely be assumed to vanish. The state equation

$$p = n \, R \, T / V \tag{7}$$

provides the pressure *p* of the ideal gas in terms of its state variables. In the above equations, *R* indicates the universal gas constant, while $c_V = c_V(T)$ is the molar specific heat of the gas at constant volume. The dependence of $c_V$ on *T* plays no role in what follows and can safely be ignored. An ideal gas is completely defined by Eqs. (6) and (7).

From Eq. (7) we calculate that

$$dW = -p \, dV = -n \frac{R \, T}{V} dV, \tag{8}$$

and hence from Eqs. (5) and (6) we obtain

$$dS = \frac{\partial S}{\partial T} dT + \frac{\partial S}{\partial V} dV = n \frac{c_V}{T} dT + n \frac{R}{V} dV. \tag{9}$$

As its derivation should make clear, this equation applies to any two adjacent states of the ideal gas that can be joined by a reversible process.

To determine the entropy function of an ideal gas, the classical approach applies Eq. (9) to a reversible constant-volume process ($dV = 0$) from state (*T*, *V*) to state (*T*+d*T*, *V*). It is thus deduced that

$$\frac{\partial S}{\partial T} = n \frac{c_V}{T}. \tag{10}$$

Similarly, by applying Eq. (9) to a reversible isothermal process ($dT = 0$) from state (*T*, *V*) to state (*T*, *V*+d*V*), the same approach obtains:

$$\frac{\partial S}{\partial V} = n \frac{R}{V}. \qquad (11)$$

By integrating Eqs. (10) and (11), it is finally concluded that to within an arbitrary constant, the entropy of an ideal gas must be given by

$$S = S(T,V) = n\, c_V \ln T + n\, R \ln V. \qquad (12)$$

(Here and in the following, we use the common practice of abbreviating ln $T$ and ln $V$ for, respectively, ln $T$/1[$T$] and ln $V$/1[$V$], where [$T$] and [$V$] are the units we have decided to use for $T$ and $V$.)

The above derivation of Eq. (12) is standard. However, it is not consistent with the physical definition of heat recalled in the previous section nor with the second law of thermodynamics. To show why, we observe that the isothermal condition d$T = 0$ used to obtain Eq. (11) must refer to a reversible process in order for Eq. (9) to apply. In such a process, no heat exchange can occur, since the gas must constantly be in thermal equilibrium with its surroundings (see, however, Section 5 for *pseudo-isothermal,* reversible, heat absorption processes). Thus, setting d$T = 0$ in Eq. (9) implies d$Q = 0$, since the system is in thermodynamic equilibrium. It follows from Eq. (3) that, in the same situation, d$S = 0$. This means that Eq. (9)$_1$, when applied to a reversible isothermal expansion from ($T$, $V$) to ($T$, $V$+d$V$), leads to the conclusion that the derivative $\partial S/\partial V$ should vanish. In other words, the entropy of the ideal gas cannot depend on $V$, in contrast to what is stated in Eq. (11).

On the basis of this result, we arrive at the conclusion that the expression for the entropy of the ideal gas must be given by

$$S = S(T) = n\, c_V \ln T. \qquad (13)$$

This equation is a correction to Eq. (12), and is consistent with the fact that no transfer of heat can occur under isothermal conditions, as implied by the thermodynamic definition of heat recalled at the beginning of Section 1.

The way in which we reached result (13), however, might appear too formal to allow tradition to be overthrown, particularly in view of the role that Eq. (12) has historically played in the statistical interpretation of entropy. For this reason, in the two sections that follow, we will derive the same result using a different analysis. Although lengthier, this new analysis has the advantage of providing new insight into the notion of entropy. We also derive a general equation to calculate the entropy of any system once the internal energy function of the system has been assigned.

## 3. Thermal and non-thermal components of the internal energy of a system

Let us refer to a generic physical system, and let $\boldsymbol{\xi} \equiv \{\xi_1, \xi_2, ..., \xi_m\}$ be the set of all the system's state variables $\xi_i$ other than $T$. The pair $(\boldsymbol{\xi}, T)$ therefore defines the state of the system. The internal energy and the entropy of the system are then functions of $\boldsymbol{\xi}$ and $T$. That is,

$$U = U(\boldsymbol{\xi}, T) \tag{14}$$

and

$$S = S(\boldsymbol{\xi}, T). \tag{15}$$

In particular systems, $U$ and $S$ may depend on appropriate subsets of the variables $\boldsymbol{\xi}$, and the subset may be different for $U$ and $S$. However, as will be shown in this section, $S$ cannot depend on state variables that $U$ does not depend upon.

If $\boldsymbol{F} \equiv \{F_1, F_2, ..., F_m\}$ denotes the (generalized) force conjugated to the variables $\boldsymbol{\xi}$, the work done on the system for infinitesimal changes in the variables $\boldsymbol{\xi}$ is given by

$$dW = \boldsymbol{F} \, d\boldsymbol{\xi} = F_1 d\xi_1 + F_2 d\xi_2 + ... + F_m d\xi_m. \tag{16}$$

In the particular case of an ideal gas, we have that $\boldsymbol{\xi} \equiv V$ and $\boldsymbol{F} \equiv -p$, which reduces Eq. (16) to Eq. (8).

A *simple heating* or *simple cooling* process is a uniform temperature process in which all the variables $\boldsymbol{\xi}$ are kept constant ($dW=0$ at any time of the process) while the system's temperature increases ($\dot{T} > 0$) or decreases ($\dot{T} < 0$), respectively. For these processes, the first law of thermodynamics (5) is reduced to

$$dU = dQ. \tag{17}$$

In this case, the heat supplied or lost by the system is equal to the internal energy change of the system.

We define the *intrinsic heat* of a system as the heat $\Gamma$ that the system would release in a simple cooling process from a given temperature $T$ to absolute zero (0 K). Since the *heat released* by the system is opposite to the *heat absorbed*, $dQ$, we have that

$$\Gamma = \Gamma(\boldsymbol{\xi}, T) = -\int_T^0 dQ \big|_{\boldsymbol{\xi}=\text{const}} = \int_0^T dQ \big|_{\boldsymbol{\xi}=\text{const}}. \tag{18}$$

From this definition and from Eq. (17), we conclude that:

$$\Gamma = \Gamma(\xi, T) = \int_0^T dU \Big|_{\xi=\text{const}} = \int_0^T \frac{\partial U(\xi, T)}{\partial T} dT . \qquad (19)$$

This equation shows that $\Gamma = \Gamma(\xi, T)$ is a state function, since such is the internal energy.

By differentiating Eq. (19) at constant $\xi$, we obtain:

$$d\Gamma \big|_{\xi=\text{const}} = \frac{\partial U}{\partial T} dT . \qquad (20)$$

That is

$$\frac{\partial \Gamma}{\partial T} = \frac{\partial U}{\partial T} . \qquad (21)$$

The last equation implies that

$$U(\xi, T) = \Gamma(\xi, T) + f(\xi) , \qquad (22)$$

where $f(\xi)$ is a differentiable function of $\xi$. On the other hand, from the definition of $\Gamma$ or, equivalently, from Eq. (18), we can infer that

$$\Gamma(\xi, 0) = 0 \text{ for every } \xi . \qquad (23)$$

From this equation and Eq. (22), it follows that

$$f(\xi) \equiv U(\xi, 0) . \qquad (24)$$

Hence, from Eqs. (22) and (24), we finally conclude that

$$\Gamma(\xi, T) = U(\xi, T) - U(\xi, 0) . \qquad (25)$$

This equation provides an alternative to Eq. (19), and is a handy way of calculating the intrinsic heat once the internal energy function is known.

As is apparent from Eqs. (22) and (24), the function $U(\xi, 0)$ represents the part of the system's internal energy in excess of the intrinsic heat $\Gamma$. Since the function $U(\xi, 0)$ does not depend on temperature, it will henceforth be referred to as the *non-thermal* component of the internal energy and denoted as $U_w$. That is:

$$U_w = U_w(\xi) = U(\xi, 0) . \qquad (26)$$

With this notation, the internal energy of any system can be generally expressed as

$$U(\pmb{\xi}, T) = \Gamma(\pmb{\xi}, T) + U_\mathrm{w}(\pmb{\xi}) \,. \tag{27}$$

This equation shows that the internal energy of a system always consists of two components: a thermal component or intrinsic heat, $\Gamma$, and a non-thermal component, $U_\mathrm{w}$, which is independent of $T$. Both components are state functions, and they are both determined by the system's internal energy function, as can be seen from Eqs. (25) and (26).

### 4. Dependence of entropy on internal energy

A change in the value of the variables $\pmb{\xi}$ makes the system absorb or supply work, according to Eq. (16). In a real process, part of this work is dissipated as heat, which may reduce the heat that the system needs to take from the surroundings. Let $\mathrm{d}\Gamma$ be the change in $\Gamma$ in a process from state ($\pmb{\xi}$, $T$) to state ($\pmb{\xi}+\mathrm{d}\pmb{\xi}$, $T+\mathrm{d}T$). From the definition of $\Gamma$, it immediately follows that $\mathrm{d}\Gamma$ measures the difference in the heat content of the system in these two states. Moreover, since $\Gamma$ is a state function, the quantity $\mathrm{d}\Gamma$ is independent of the process joining these states. Thus, any process from state ($\pmb{\xi}$, $T$) to state ($\pmb{\xi}+\mathrm{d}\pmb{\xi}$, $T+\mathrm{d}T$) must supply the system with the amount of heat $\mathrm{d}\Gamma$. If $\mathrm{d}\Gamma < 0$, the system releases heat.

By inserting Eq. (27) into Eq. (1), we obtain

$$\mathrm{d}\Gamma(\pmb{\xi}, T) + \mathrm{d}U_\mathrm{w}(\pmb{\xi}) = \mathrm{d}W + \mathrm{d}Q \,. \tag{28}$$

The non-thermal component $U_\mathrm{w}$ of the internal energy does not depend on the heat absorbed by the system. Its change $\mathrm{d}U_\mathrm{w}$ must therefore be produced by $\mathrm{d}W$. In contrast, both $\mathrm{d}W$ and $\mathrm{d}Q$ can contribute to $\mathrm{d}\Gamma$. This implies that depending on the process, the amount of heat $\mathrm{d}Q$ that the system takes from its surroundings may be less than $\mathrm{d}\Gamma$.

More specifically, when the process is irreversible, part of the work done on the system dissipates as heat, which, if absorbed by the system, makes $\mathrm{d}W > \mathrm{d}U_\mathrm{w}$. Due to the heat produced by the dissipated work, the amount of heat $\mathrm{d}Q$ that the system needs to take from the surroundings is reduced, which makes $\mathrm{d}Q < \mathrm{d}\Gamma$. Thus, the occurrence of dissipation in a process leads to an increase in the work $\mathrm{d}W$ that must be done on the system and a reduction in the heat that the same system absorbs from its surroundings.

In the absence of dissipation, i.e. in reversible processes, no work is turned into heat. In this case, the increase in the intrinsic heat of the system is entirely due to $\mathrm{d}Q$, and we have

$$\mathrm{d}Q = \mathrm{d}\Gamma \,. \tag{29}$$

We can conclude that the quantity d$\Gamma$ represents the maximum amount of heat that the system can ever absorb from its surroundings when it undergoes a process from ($\boldsymbol{\xi}$, $T$) to ($\boldsymbol{\xi}$+d$\boldsymbol{\xi}$, $T$+d$T$). This means that the relation

$$dQ \leq d\Gamma \qquad (30)$$

must apply to every process, the equality sign being valid if the process is reversible.

A corollary of this result is that the difference $\Delta\Gamma$ in the values that the intrinsic heat attains at any two states represents the maximum amount of heat that the system can ever absorb in a process joining the two states. This fact justifies referring to $\Delta\Gamma$ as the *maximum absorbable heat*.

To better appreciate the import of limitation (30), we can refer to any two states of the system, say state $A$ and state $B$, which are not necessarily in the same infinitesimal neighborhood. The maximum absorbable heat in a process from $A$ to $B$ is then given by $\Delta\Gamma_{AB} = \Gamma_B - \Gamma_A$, where $\Gamma_A$ and $\Gamma_B$ denote the values of the intrinsic heat at $A$ and $B$, respectively. If $\Delta\Gamma_{AB}$ is greater than zero, it represents the maximum amount of heat that the system can ever absorb from the surroundings in a process from $A$ to $B$.

Assuming $\Delta\Gamma_{AB} > 0$, the analogous calculations for the inverse process, i.e. from $B$ to $A$, yields $\Delta\Gamma_{BA} = \Gamma_A - \Gamma_B = -\Delta\Gamma_{AB} < 0$. The negative value of $\Delta\Gamma_{BA}$ indicates that the system does actually supply heat to the surroundings. From relation (30), we infer that if a process is dissipative, the heat that the system supplies to the surroundings in the inverse process is greater than the maximum amount of heat that the system can absorb in the direct process. Thus, a cycle from $A$ to $B$ and then back to $A$ can never result in an overall net absorption of heat from the surroundings. (Of course, this does not exclude the possibility that the system may absorb heat from one part of the surroundings and return even more heat to another part, as done by refrigerating machines.)

No similar limitation applies to the maximum amount of heat that a system can supply to the surroundings. The reason is that relation (30) sets no limit on the maximum negative value that d$Q$ can assume in a process. Hence, provided that we have enough energy to dissipate, we can supply heat to the surroundings at any rate and in any amount, possibly by repeating the same dissipative process many times—a phenomenon that Count Rumford observed more than two centuries ago in his celebrated cannon-boring experiments.

A crucial point that emerges from the above analysis is that in any infinitesimal part of a reversible process, the amount of heat d$Q$ absorbed by the system must meet the conditions of both Eq. (3) and Eq. (29). This fact implies that

$$\mathrm{d}\Gamma = T\,\mathrm{d}S, \tag{31}$$

which shows how $\Gamma$ and $S$ are related to each other. Since both $\Gamma$ and $S$ are state functions, this equation applies regardless of the process, although it was obtained by referring to a reversible process.

Equation (31) can also be obtained from the classical thermodynamic definition of entropy:

$$S = S(\boldsymbol{\xi}, T) = \int_{T=0\,\mathrm{K}}^{(\boldsymbol{\xi},T)} \left.\frac{\mathrm{d}Q}{T}\right|_{\mathrm{rev}}, \tag{32}$$

which applies to any reversible process that starts from a state at the temperature of absolute zero and ends at state ($\boldsymbol{\xi}$, $T$). (The value assumed by $\boldsymbol{\xi}$ at $T = 0$ K is immaterial, because the third law of thermodynamics states that at absolute zero, the entropy of a system assumes a constant value, irrespective of the value of $\boldsymbol{\xi}$.) To prove Eq. (31), let us apply Eq. (32) to a process in which all the state variables other than $T$ are kept constant at their final value $\boldsymbol{\xi}$. This is a simple heating process, as defined in Section 3. For this process, by differentiating Eq. (32) we can infer that

$$T\,\mathrm{d}S = \mathrm{d}Q|_{\boldsymbol{\xi}=\mathrm{const}}. \tag{33}$$

Similarly, by differentiating Eq. (18) we get:

$$\mathrm{d}\Gamma = \mathrm{d}Q|_{\boldsymbol{\xi}=\mathrm{const}}. \tag{34}$$

By comparing this equation to Eq. (33), we obtain Eq. (31). This completes the proof.

Let us now write Eq. (31) in the form:

$$\mathrm{d}S = \frac{\mathrm{d}\Gamma}{T} = \mathrm{d}\!\left(\frac{\Gamma}{T}\right) + \frac{1}{T^2}\Gamma\,\mathrm{d}T. \tag{35}$$

Upon integration, this equation yields:

$$S = S(\boldsymbol{\xi}, T) = \frac{\Gamma(\boldsymbol{\xi}, T)}{T} + \int \frac{\Gamma(\boldsymbol{\xi}, T)}{T^2}\,\mathrm{d}T + S_0, \tag{36}$$

where $S_0$ represents an arbitrary integration constant. By using Eq. (27), we can express Eq. (36) as:

$$S = S(\xi, T) = \frac{U(\xi, T) - U_w(\xi)}{T} + \int \frac{U(\xi, T)}{T^2} \, dT - \int \frac{U_w(\xi)}{T^2} \, dT + S_0 \,. \tag{37}$$

This relation is quite general, and provides the means of determining the entropy of a system from its internal energy.

In the case of an ideal gas, we have that $U=U(T)$, according to Eq. (6). From Eq. (26), we can then infer that $U_w \equiv 0$ in the ideal gases, since at a temperature of absolute zero the internal energy of an ideal gas is taken to be equal to zero. By introducing Eq. (6) into Eq. (37), we then obtain

$$S = S(T) = n\, c_v + n\, c_v \int \frac{1}{T} \, dT + S_0 \,. \tag{38}$$

From this, by setting $S_0 = -n\, c_V$, we finally get

$$S = S(T) = n\, c_v \ln T \,, \tag{39}$$

which provides an alternative proof of Eq. (13).

An important consequence of Eq. (37) is that the entropy of a system cannot depend on state variables on which the internal energy is not dependent. This simple rule would have sufficed to exclude from the outset the validity of Eq. (12) as a possible expression for the entropy of a gas possessing the internal energy (6).

Observe, however, that the same rule does not exclude the possibility that the entropy of a system may depend on only a few of the state variables which appear in the internal energy function. The reason for this is that some of these variables may drop out of Eq. (37) after simplification.

## 5. Isothermal and pseudo-isothermal processes

The impossibility of heat transfer between two bodies at the same temperature means that any isothermal process of a system in thermal equilibrium with its surroundings is also adiabatic. For these processes, therefore, $dQ = 0$ and Eq. (1) becomes

$$dU = dW \,, \tag{40}$$

which coincides with the expression of the first law under adiabatic conditions.

On the other hand, any isothermal process of an ideal gas occurs at $dU = 0$, as is apparent from Eq. (6). When applied to an ideal gas, therefore, Eq. (40) is reduced to

$$dW = 0 \,. \tag{41}$$

This means that an ideal gas cannot produce or absorb work while undergoing an isothermal volume change in thermal equilibrium with its surroundings; or, equivalently, any reversible isothermal process of an ideal gas must be workless.

Incidentally, it should be noted that this result is a consequence of the fact that the internal energy of the ideal gas is a function only of $T$. The expression for the entropy of the gas plays no role here, as is apparent from the analysis given above.

The restriction imposed by Eq. (41) may appear to contradict the experimental evidence that a gas can produce work while expanding isothermally and at the same time absorbing heat from the surroundings. The gas in question may well be within the range of temperature and pressure where it behaves as an ideal gas, and the process may be executed reversibly to any desired degree of accuracy. The point to be noted here, however, is that these processes are not strictly isothermal. They are a sequence of small adiabatic expansions in which the gas cools, its internal energy decreases and work is done, followed by constant volume heat exchanges in which the gas worklessly regains its original temperature by exchanging heat with its surroundings.

This sequence of adiabatic expansions and constant-volume heat exchanges becomes reversible if performed sufficiently slowly and in small steps. Moreover, by regulating the amplitude of the steps, the temperature changes at each step can be made as small as desired. Thus, although the temperature of the gas over the whole process can be maintained at a nearly constant value, to any desired degree of accuracy, these processes are not strictly isothermal, because they are made up of non-isothermal steps. In particular, at no time in the process is heat exchanged at $dT = 0$. In what follows, a process of this kind is referred to as a *pseudo-isothermal process*. Figure (1) provides a sketch of a few steps in these processes.

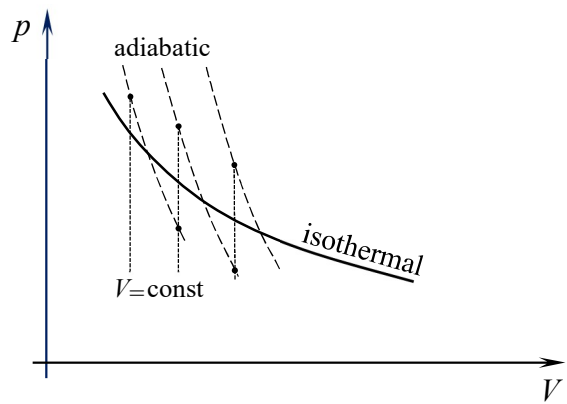

**Fig. (1)**. Adiabatic (work-producing) and constant-volume (heat exchanging) steps making up a *pseudo-isothermal process*. Although there are no points in the process where the temperature remains constant, it is possible to maintain a pseudo-isothermal process as close to an isothermal curve as desired.

The work $W$ done by an ideal gas in a reversible, pseudo-isothermal process from volume $V_1$ to volume $V_2$ is obtained by integrating Eq. (8) along the adiabatic parts of the process, since the constant-volume parts produce no work. In this integration, we can treat the gas temperature $T$ as a constant, because the process is constructed so as to keep the temperature fluctuations small with respect to $T$. We therefore have:

$$W = \int_{V_1}^{V_2} dW = -n \int_{V_1}^{V_2} \frac{RT}{V} dV = n R T \ln \frac{V_1}{V_2} . \qquad (42)$$

On the other hand, in a pseudo-isothermal process, each complete step of adiabatic expansion followed by a constant volume heat exchange makes the final temperature of the gas equal to the initial one. Since $T$ is the only independent variable of the internal energy of an ideal gas, it follows that a pseudo-isothermal process and any of its complete steps do not produce any final change in the gas internal energy. Therefore, by integrating Eq. (1) between the initial and final state of a complete pseudo-isothermal process, we can conclude that the heat $Q$ absorbed by the gas in such a process is given by

$$Q = -W . \qquad (43)$$

This result is valid regardless of the number of the steps of adiabatic expansion with subsequent constant volume heat exchange that make up the process. It should be noted, however, that Eq. (43) is an integral result. As such, it refers to an entire process consisting of an integer number of complete steps of adiabatic expansion and constant volume heat exchange, as defined above. The same result in no way means that $dQ = dW$ throughout a pseudo-isothermal process, as would be true if Eq. (43) applied at any time of the process. The opposite is true. In a pseudo-isothermal process, we have that $dQ \neq 0$ and $dW = 0$ in the constant volume parts of the process, while $dQ = 0$ and $dW \neq 0$ in the adiabatic parts. However, the complete step or process makes $Q = -W$, as stated by Eq (43).

The pseudo-isothermal process is not isentropic. However, the same process and its steps of adiabatic expansion with the associated constant volume heat exchange do not cause any final change in the entropy of the ideal gas. This fact is the direct consequence of Eq. (13) and the equality of the initial and final temperature that the process produces.

Equations (42) and (43) coincide with the classical thermodynamic expressions for the work and heat absorbed by an ideal gas in a reversible isothermal process (cf. e.g. [1], p. 57 and [17], p. 23). However, the classical theory assumes that the same equations are valid for strict isothermal processes, which would imply that the gas can absorb heat from its surroundings while remaining in thermal equilibrium with them at a constant temperature.

Though consistent with the first law, such behavior would not be compatible with the concept of heat recalled at the beginning of Section 1. The pseudo-isothermal processes introduced above are free from this contradiction.

If properly executed, a pseudo-isothermal process can approximate a reversible isothermal process to any degree of accuracy. Moreover, the work and heat absorbed by an ideal gas in a pseudo-isothermal process are given by the same expressions as those traditionally used for an isothermal process of an ideal gas. Hence, with the proviso that what is classically considered an isothermal process is actually a pseudo-isothermal process, most of the classical results concerning the work and heat absorbed by an ideal gas under isothermal conditions remain valid.

**6. Consistency with the third law**

By integrating Eq. (1) between the initial and final states of a process, we can write:

$$\Delta U = \Delta W + \Delta Q, \tag{44}$$

where $\Delta U = U_2 - U_1$ denotes the difference between the final value $U_2$ and the initial value $U_1$ of the internal energy, while $\Delta W$ and $\Delta Q$ are the work done on the system and the heat absorbed by it in the process, respectively.

Consider then the free expansion of a gas in a vacuum, starting from volume $V_1$ and ending at volume $V_2$. The work $\Delta W$ vanishes, since there is no external pressure. Assuming that the expansion occurs in a thermally isolated container, the heat absorbed by the gas is also zero. For a gas undergoing this free expansion process, we therefore have $\Delta W = \Delta Q = 0$, and Eq. (44) becomes:

$$\Delta U = 0. \tag{45}$$

That is, the adiabatic free expansion of a gas does not change the internal energy of the gas. If the expanding gas is ideal, the additional condition $\Delta T = T_2 - T_1 = 0$ applies, as immediately follows from Eq. (6), $T_1$ and $T_2$ being the initial and the final temperatures of the gas. Hence, the adiabatic free expansion of an ideal gas is also isothermal.

However, if Eq. (12) were valid, the same process would produce a change in the entropy of the ideal gas, given by

$$\Delta S = n R \ln \frac{V_2}{V_1}. \tag{46}$$

Thus, by increasing the expansion ratio $V_2/V_1$ appropriately, we could make the entropy of any given mass of gas as large as we wish, without expending any heat or work, while leaving the internal energy of the gas unaltered. The third law excludes this possibility, as shown below.

Let us for a moment refer to an ideal gas undergoing a reversible simple-cooling-process from a temperature $T$ to 0 K at a given constant volume $V$. For this process, the first law is reduced to Eq. (17), so that by integrating Eq. (3) we have:

$$\int_T^{0\,\mathrm{K}} \mathrm{d}S = \int_T^{0\,\mathrm{K}} \frac{\mathrm{d}Q}{T} = \int_T^{0\,\mathrm{K}} \frac{\mathrm{d}U}{T}. \tag{47}$$

Since in an ideal gas we have that $U=U(T)$, the value of the entropy integral appearing in this equation does not depend on the volume at which the simple cooling process takes place. Hence, by applying Eq. (47) to two different simple cooling processes at volumes $V_1$ and $V_2$, respectively, we can infer that

$$\int_T^{0\,\mathrm{K}} \mathrm{d}S\Big|_{V=V_1} = \int_T^{0\,\mathrm{K}} \mathrm{d}S\Big|_{V=V_2}. \tag{48}$$

From the third law of thermodynamics, we know that the entropy of a system in thermodynamic equilibrium approaches a constant value as its temperature approaches absolute zero, regardless of the values of the state variables of the system other than temperature. Thus, by denoting by $S_0$ the entropy at absolute zero, we can write Eq. (48) as

$$S_0 - S_1 = S_0 - S_2. \tag{49}$$

That is,

$$S_1 = S_2, \tag{50}$$

where $S_1$ and $S_2$ are the initial values of the gas entropy at the beginning of the simple cooling processes at volumes $V_1$ and $V_2$, respectively. The classical expression for the entropy of an ideal gas is incompatible with this result. As is apparent from Eq. (12), this implies that for any given $T$, we have that $S_1 \neq S_2$ for $V_1 \neq V_2$, in contrast to Eq. (50).

Let us now return to the free expansion of an ideal gas from volume $V_1$ to volume $V_2$ considered at the beginning of this section. The temperature $T$ at the beginning and end of the expansion is the same, since as observed above, the free expansion of the gas is an isothermal process. For the amount of gas considered here, we can always imagine two simple cooling processes from $T$ to 0 K, taking place at volume $V_1$ and volume $V_2$, respectively. From Eq. (50), we know that the initial values of the entropy of the gas in these simple cooling processes must be the same. This means that in the free expansion process from $V_1$ to $V_2$, the initial and final values of the entropy of the gas must also be the same. This is valid irrespective of the value of $V_2$, provided that it is larger than $V_1$. We can therefore conclude that the free expansion of an ideal gas in a vacuum is isentropic. This result is consistent with Eq. (13) for the ideal gas entropy, but not with the classical Eq. (12).

We finally observe that since it is both adiabatic ($dQ = 0$) and isentropic ($dS = 0$), the free expansion of an ideal gas meets condition (3), and is therefore thermodynamically reversible. Of course, in order to undo a free expansion in a reversible way, a reversible, workless volume contraction is needed. One way to realize a workless volume contraction is to cool the gas to as near to absolute zero as possible, thus making the gas pressure drop to zero to any desired degree of accuracy. The volume reduction of the gas can then be achieved worklessly, say by pushing a frictionless piston. Since the entropy is constant at 0 K, this volume reduction can be made both adiabatic and isentropic—and thus reversible—to any desired degree of accuracy.

In real gases, the process of volume reduction occurs spontaneously above a temperature of absolute zero. At low temperatures, the gas liquefies, thus making its volume small enough to allow the initial volume of the gas to be recovered via a subsequent reversible heating process.

In general, for any gas, there is no conceptual difficulty in executing a reversible cooling process that brings the expanded gas close to 0 K and a heating process that brings it back to the original temperature $T$. For an ideal gas in particular, simple cooling/heating processes performed at the volume occupied by the gas after/before the free expansion process under consideration can serve this purpose. Since the specific heat of the gas does not depend on volume, the heat produced in the simple cooling process equals the heat absorbed by the gas in the subsequent simple heating process. A full reversible recovery of the initial state of the gas can therefore be achieved.

## 7. Statistical entropy

In the middle of the 1860s, James Clerk Maxwell determined the expected distribution of the velocities of gas molecules from the mechanics of their collisions. The ensuing kinetic theory of gases could explain through the laws of mechanics many of the thermodynamic properties of a gas, such as its specific heats, temperature, pressure, and viscosity. Entropy was not even mentioned. The spectacular success of Maxwell's theory initiated intense research activity with the aim of deducing the whole of thermodynamics from mechanics. The efforts in this direction had of necessity to take a statistical approach, and eventually led to what is now widely referred to as the Boltzmann (or Boltzmann-Planck) entropy equation:

$$S = S(U,V,N) = k \ln \Omega. \tag{51}$$

This equation defines the *statistical entropy* of an ideal gas (cf., e.g., [17, p.61]). In statistical thermodynamics, the macrostate of an ideal gas is defined by its internal energy $U$, volume $V$, and number of particles $N$. The gas entropy, $S$, in a given macrostate is then a function of the number $\Omega = \Omega(U, V, N)$ of microstates (i.e., the number of all possible collections of position and momentum of the gas particles) that are compatible with the macrostate. The quantity $k$ appearing in Eq. (51) is the Boltzmann constant.

A modified version of Eq. (51) was proposed by J. Willard Gibbs to calculate the entropy in more general systems, when the microstates corresponding to a given macrostate of the system have different probabilities of occurring:

$$S = -k \sum_i p_i \ln p_i. \tag{52}$$

Here, the quantity $p_i$ is the probability of finding the system in the $i$-th microstate.

It is widely acknowledged that Eq. (51), and thus its generalization (52), is, in fact, an assumption. It is introduced in statistical mechanics to give the entropy a microscopic interpretation. The main argument in support of this assumption is that when it is applied to an ideal gas, it provides results that are consistent with those obtained from the classical Eq. (12) (cf. e.g. [17] and [1]). However, we showed in Section 2 that Eq. (12) is not the correct expression for the ideal gas entropy, as it requires the entropy to depend on the gas volume, and this is inconsistent with the internal energy of the ideal gas, which is independent of $V$. The correct expression for the entropy of the ideal gas was obtained in the same section and is volume-independent, see Eq. (13). In contrast, Eqs. (51) and (52) produce an entropy that depends on the volume because the number of different positions of the gas particles, and so $\Omega$, depends on $V$.

This assertion is supported by the fact that the expression for the internal energy of an ideal gas—that is, the total kinetic energy of its particles—is independent of the $\Omega$ number of microstates. This independence is a direct consequence of the first law of thermodynamics, as the energy of a closed system in a given macroscopic state cannot depend on the number of available microstates, and even less on the probability that the system is in one of them. Entropy must follow the same logic since it is connected to the internal energy of the system via Eq. (37). So, at variance with Eq. (51), the entropy of an ideal gas cannot depend on $\Omega$, which is consistent with the conclusion we reached at the end of Section 4.

It should also be observed that the definition of statistical entropy is still object of some debate. For an interesting review of its origin and possible generalizations the reader is referred to [18] and [19]. Whatever the case, statistical entropy appears to be a notion organically linked to the statistical mechanics view of physics and fundamentally different from the entropy used in macroscopic thermodynamics.

**8. Remarks on the experimental determination of entropy.**

The most direct experiment to determine the entropy of a substance is based on Eq. (4). In this experiment, the substance's temperature and the amount of heat it absorbed are measured during a reversible heating or cooling process from the initial state A to the final state B (henceforth simply referred to as the heating process). The obtained data are used to calculate the integral entering Eq. (4) and, hence, the entropy difference $\Delta S = S_B - S_A$ between the two considered states.

When operating in a range of temperature and pressure in which the gas can, for all practical purposes, be treated as ideal, the above procedure is often performed at constant pressure. The majority of the data on gas entropy reported in standard chemical tables available in the literature comes from constant-pressure gas heating processes. This is also the case of the experiments considered in [20] to assess the validity of the entropy constants of monoatomic gases derived from statistical thermodynamics using the Sackur-Tetrode equation. Correct as they are, the entropy values thus determined cannot be used to assess whether or not the entropy of an ideal gas depends on volume. The proof is the following.

Let's refer to $n$ moles of an ideal gas undergoing a process at a constant pressure $p$. For each infinitesimal step of the process, the volume change of the gas can be calculated from Eq. (7) to be given by

$$dV = \frac{nR}{p} dT , \qquad (53)$$

while the work done on the gas (or *by* the gas, if it is negative) is given by

$$dW = -p\, dV = -nR\, dT , \qquad (54)$$

as immediately follows from Eqs. (7), (8) and (53). In the same infinitesimal step, the heat $dQ$ that the gas exchanges with is surroundings is

$$dQ = dU - dW = n\,(c_V + R)\, dT, \qquad (55)$$

as follows from Eqs. (1), (6) and (54). The entropy change of the gas due to a heating process at constant $p$ is finally obtained by integrating Eq. (4) once Eq. (55) is inserted into it. That is:

$$\Delta S = S_B - S_A = \int_A^B \frac{dQ}{T}\bigg|_{rev} = n(c_v + R)[\ln T_B - \ln T_A] = n(c_v + R)\ln \frac{T_B}{T_A}. \tag{56}$$

As apparent from this equation, the considered entropy change is the same regardless of whether the entropy of the gas depends on volume or not. Therefore, with a constant pressure heating process, it is not possible to prove or disprove that the entropy of a gas depends on its volume.

## 9. Conclusion

Various physical theories have exploited Eqs. (51) and (52) to introduce the notion of entropy; two notable instances are quantum mechanics and information theory. Quantum mechanics derived the so-called von Newman entropy from Eq. (52), while in information theory, Eq. (52) gave rise to what is known as the Shannon entropy. Several other entropies are introduced in other branches of physics to put them in the broader framework of the second law of thermodynamics. Some approaches consider entropy to be a measure of the number of microstates available, while others interpret it as a measure of the probability that a system will be in a specific microstate. Certain approaches view entropy as a measure of microscopic order/disorder, and still others regard it as a measure of our lack of information about the microscopic state of the system. These extended entropies are all descended from Eq. (51). None of them is legitimate in regard to the second law of thermodynamics since, as repeatedly stated here, Eq. (51) does not represent a correct expression of the thermodynamic entropy.

The above conclusion in no way means that quantities such as $\Omega$ and $p_i$ are inadequate to describe the increasing state of disorder that many granular systems, such as those made of small solid particles, may exhibit when acted upon by external forces. A typical example is a half-full box of red and white balls, initially arranged in a well-ordered red and white layer. Shaking the box mixes the red and white balls, causing them to assume one of a myriad of disordered configurations corresponding to a higher value of $\Omega$ and possessing a greater probability of occurring than the initial configuration. Although this phenomenon is hardly invertible, it is not irreversible in the thermodynamic sense, since heat and temperature play a minimal role, if any, in the shaking process.

The motion of the spheres considered above is highly sensitive to small changes in their initial conditions. Small perturbations in the current trajectories of the balls lead to tremendous changes in their future positions, resulting in disordered mixing of the balls. Phenomena of this kind belong to chaos theory, and are governed by mechanics or simple kinematics. The quantities $\Omega$ and $p_i$ represent, respectively, the number of possible configurations of the balls and the probability that a specific arrangement of the balls will occur. Although both quantities reach their maximum in the most disordered configuration, this has nothing to do with the upper limit on the system's ability to absorb heat set by thermodynamic entropy and the second law of thermodynamics.

Incidentally, this conclusion helps settle the often-made but controversial claim that the second law drives the universe to chaos by preventing the entropy of a closed system from decreasing. This claim does not appear to be supported by the universe.